# Thermal Property Microscopy with Compressive Sensing Frequency-Domain Thermoreflectance

Haobo Yang[1,2], Zhenguo Zhu[3], Zhongnan Xie[4], Jinhong Du[3], Shuo Bai[3], Hong Guo[4,5], Te-Huan Liu[1], Ronggui Yang[1,6*], Xin Qian[1*]

[1] School of Energy and Power Engineering, Huazhong University of Science and Technology, Wuhan, Hubei 430074, China

[2] State Key Laboratory of Coal Combustion, Huazhong University of Science and Technology, Wuhan, Hubei 430074, China

[3] Shenyang National Laboratory for Materials Science, Institute of Metal Research, Chinese Academy of Sciences, Shenyang 110016, China

[4] State Key Laboratory of Nonferrous Metals and Processes, China General Research Institute for Nonferrous Metals Group Co., Ltd., Beijing, 100088, China

[5] China GRINM Group Engineering Institute Co., Ltd., Beijing, 101400, China

[6] College of Engineering, Peking University, Beijing 100871, China

*Corresponding emails: xinqian21@hust.edu.cn; ronggui@pku.edu.cn;

**ABSTRACT**

Spatial mapping of thermal properties is critical for unveiling the structure-property relation of materials, heterogeneous interfaces, and devices. These property images can also serve as datasets for training artificial intelligence models for material discoveries and optimization. Here we introduce a high-throughput thermal property imaging method called compressive sensing frequency domain thermoreflectance (CS-FDTR), which can robustly profile thermal property distributions with micrometer resolutions while requiring only a random subset of pixels being experimentally measured. The high-resolution thermal property image is reconstructed from the raw down-sampled data through $\mathcal{L}_1$-regularized minimization. The high-throughput imaging capability of CS-FDTR is validated using the following cases: (a) the thermal conductance of a patterned heterogeneous interface, (b) thermal conductivity variations of an annealed pyrolytic graphite sample, and (c) the sharp change in thermal conductivity across a vertical aluminum/graphite interface. With less than half of the pixels being experimentally sampled, the thermal property images measured using CS-FDTR show nice agreements with the ground truth (point-by-point scanning), with a relative deviation below 15%. This work opens the possibility of high-throughput thermal property imaging without sacrificing the data quality, which is critical for materials discovery and screening.

# I. INTRODUCTION

Mapping thermal properties with microscale resolution is essential for developing advanced materials and for thermal management. For example, thermal conductivity imaging can be used to study the phonon-defect scattering mechanisms, and identify the intrinsic thermal transport properties by performing local measurements in the low-defect-concentration region. A notable example is the experimental confirmation of the ultrahigh thermal conductivity (~ 1300 W m$^{-1}$ K$^{-1}$) of boron arsenide by thermal property mapping at the sample facet [1–3]. Thermal property imaging can also be useful for optimizing the composition of multicomponent alloys. By profiling the properties across a wedge-shaped combinatorial sample, thermal property variations across the entire phase diagram can be characterized, from which the optimal composition with the target property can be identified [4–7]. Imaging thermal conductance across buried interfaces can be used for detecting weakly bonded regions [8,9], which is vital to ensure efficient heat removal from hot spots in integrated circuits. With the increasing momentum of artificial intelligence (AI) for science [10,11], thermal property images can serve as high-fidelity experimental datasets for training AI models for material discovery and device design and optimization, given that the imaging throughput can be significantly improved to generate big data.

Present techniques for imaging thermal property at the microscale include scanning thermal microscopy (SThM) [12] and pump-probe thermoreflectance techniques [13]. SThM utilizes cantilevers of an atomic force microscope (AFM) to transduce the thermal expansion of the sample with nanometer spatial resolution [14,15]. However, it requires complicated and careful fabrication of the detecting tips, and is extremely sensitive to the morphology of the sample surfaces [16]. Pump-probe thermoreflectance techniques, such as time-domain thermoreflectance (TDTR) [17] and frequency-domain thermoreflectance (FDTR) [18], use a pump beam for thermal excitation and a probe beam to detect thermal responses by measuring the local reflectance change. The advantages of pump-probe thermoreflectance techniques include minimal sample preparation, the ability to perform *in situ* measurements, and the capability of resolving thermal transport properties of buried thin films and interfaces in multi-layered structures. FDTR and TDTR have been widely used for characterizing specific heat and thermal conductivity of bulk materials, thin films, and interface thermal conductance across heterogeneous interfaces [13,17]. Both techniques can be extended to thermal property microscopy with microscale resolutions by performing local measurements and scanning over the sample surface in a point-by-point manner [19,20]. In addition to scanning imaging techniques, CCD- or IR-based wide-field microscopy [21–23] has been proposed to directly measure temperature profiles. However, reconstructing the thermal property profiles from temperature images is a complicated inverse problem, which remains challenging. To obtain thermal property images with high spatial resolutions, a large number of pixels needs to be measured, resulting in a long imaging time and a large amount of data to be processed.

In this work, we introduce a high-throughput thermal property imaging method called compressive sensing frequency domain thermoreflectance (CS-FDTR), which can robustly profile thermal property distributions with micrometer resolutions while requiring only a random subset of pixels being experimentally measured. Compressive sensing leverages the idea that a sparse signal can be reconstructed from fewer measurements than required by the Shannon-Nyquist theorem [24], which has been applied in many common imaging techniques such as magnetic resonance imaging (MRI) [25], thermoacoustic imaging [26], ultrafast spectroscopy [27], and many others [28–30]. CS-FDTR replaces the conventional point-by-point sampling with sparse random sampling, and the high-resolution image is then reconstructed from the down-sampled raw image by solving an $\mathcal{L}_1$-regularized minimization problem. This technique can simultaneously reduce the imaging time and the computational cost for data processing by a factor of 2~6, without sacrificing the accuracy of property imaging. The proposed CS-FDTR method is applied to construct the images of thermal conductance across a patterned interface, spatial variations in thermal conductivity of an annealed pyrolytic graphite (APG), and near a heterogeneous interface between aluminum and graphite. By measuring only 15% to 50% of the pixels, the reconstructed thermal property profiles in CS-FDTR show a normalized root-mean-square error below 15% compared with the ground truth images. This work pushes forward the development of high-throughput thermal property characterizations that are critical for materials discovery, screening, and optimization.

## II. EXPERIMENTAL METHODS

This section describes the general principles and workflow using CS-FDTR for thermal property imaging, details in the implementation of the FDTR system for local measurements, and the imaging reconstruction by solving a $\mathcal{L}_1$-regularized minimization problem.

### A. General principles of CS-FDTR

Compressive sensing leverages the sparse feature of realistic signals to enhance the throughput of measurements, such that the essential information embedded in the signal can be captured with fewer experiments than the Shannon-Nyquist theorem [31,32]. In the case of thermal property imaging, a realistic thermal property distribution can be represented as Fourier coefficients that capture the property variations at different length scales. The low-frequency coefficients of the image correspond to the clustering of similar properties in certain regions, while the high-frequency components capture the fluctuations or detailed variations within a few pixels. There must be a minimal length scale $l_{min}$, below which the thermal property variations are negligible. For example, the $l_{min}$ of a nanocomposite sample with heterogeneous fillers corresponds to the minimal size of the fillers. In the Fourier-transformed domain, the thermal property profile will only contain nonzero components below the spatial frequency of $1/l_{min}$, above which the components are small and physically insignificant noises. As a result, the number of nonzero components in the Fourier domain is typically less than the total number of pixels in the property image, *i.e.* the image is sparse.

Conventional sampling with equal intervals is dictated by the Shannon sampling theorem, where the sampling interval must be smaller than $l_{min}/2$ to preserve the high-frequency variations in the property profile. This is because a uniform sampling corresponds to an equally spaced series of spatial frequencies in the Fourier domain, and the information at the uncovered frequencies is lost. However, if nonuniform sampling is performed in real space, an ensemble of spatial frequency components is measured at the same time, which is called incoherent sampling [31]. According to the compressive sensing theory, much fewer measurements are required using such incoherent sampling without losing the major features, and the primary components with large amplitudes can be accurately calculated by solving a $\mathcal{L}_1$-regularized minimization problem.

The workflow of CS-FDTR is outlined in Figure 1. As shown in Figure 1(a), CS-FDTR first takes random local measurements, which can be represented by a random sampling matrix $\Pi$ of bool values, with the sampled location labeled as one and the rest labeled as zero. The raw down-sampled image of thermal properties using FDTR is shown in Figure 1(b), with lots of "blind spots" represented by the white pixels. Then, the CS-FDTR approach seeks to reconstruct the image and fill the blind spots with reasonable thermal property values. This reconstruction step is done in the Fourier-transformed domain. The task of image reconstruction is to calculate the optimal Fourier coefficients from the down-sampled data using $\mathcal{L}_1$-regularized minimization. Such reconstruction is feasible because (i) both the low- the high-spatial-frequency variations are captured by local measurements with randomly varied distances, and (ii) the primary nonzero components capturing the clustering of similar properties are concentrated in the low-frequency ranges as shown in Figure 1(c). After the Fourier coefficients are calculated from experimental data, the thermal property image is reconstructed by performing the inverse DCT transform, as shown in Figure 1(d).

Using the down-sampling and reconstruction scheme, the CS-FDTR technique can substantially increase the imaging throughput compared with point-by-point sampling. When imaging thermal properties inside a 100 μm×100 μm region, point-by-point imaging would take more than 1 hour, supposing a scanning step of 1 μm is used. The time required for each local FDTR measurement traditionally takes a few tenths of a second, which is mainly determined by the time constant of the lock-in amplifier to maintain a high signal-to-noise ratio. Processing the measurement data with $10^4$ pixels would take another 30 minutes, even with parallelized data reduction running on a workstation with a 48-core Inter Xeon Folden 6248R

CPU. Using CS-FDTR with down-sampling, both the acquisition time and data processing can be greatly reduced by a factor of 2~6, depending on the sampling fraction and the sparsity of the thermal property image. The computational cost for image reconstruction is negligible, requiring only a few seconds to solve the $\mathcal{L}_1$-regularized minimization problem. In the following Sections II.B and II.C, details of sparse local FDTR measurements and image reconstruction are discussed.

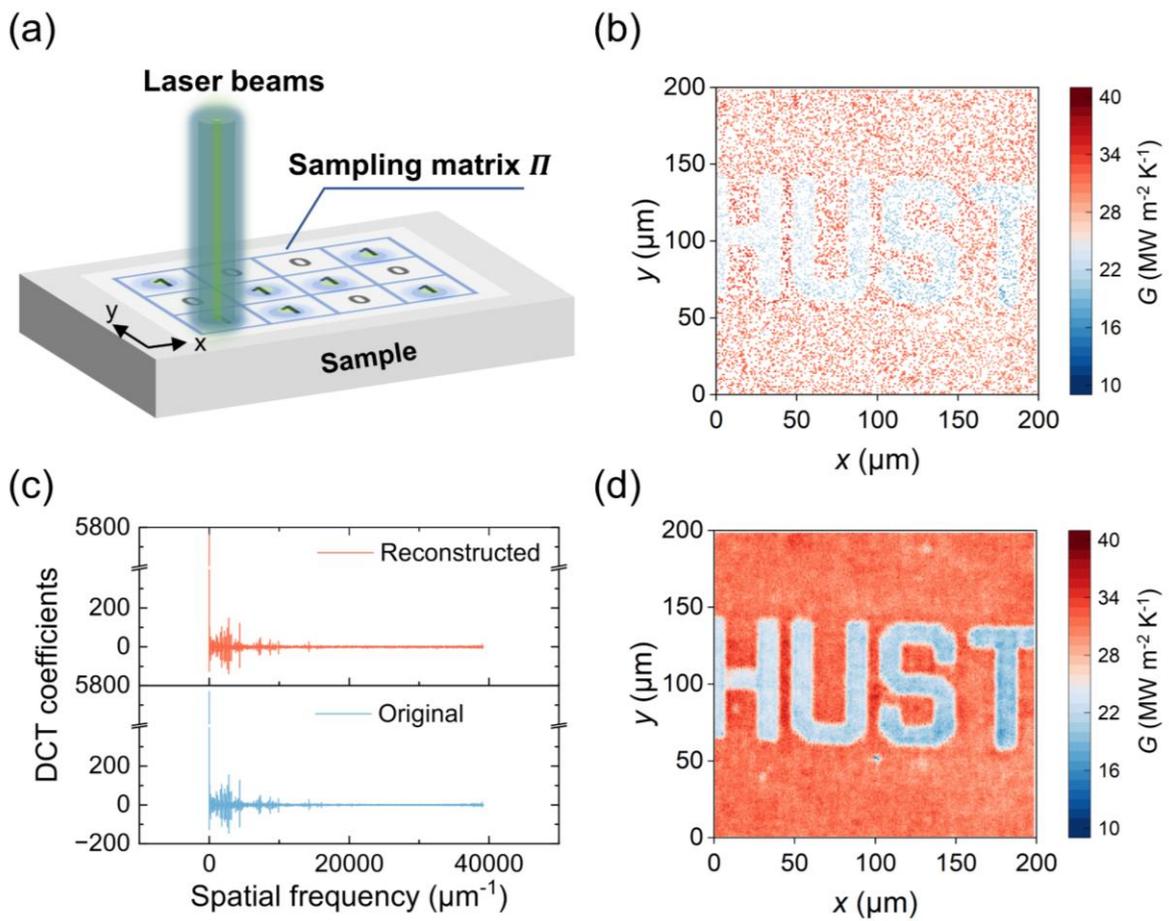

**Figure 1. The general workflow of CS-FDTR.** (a) Illustration of the down-sampling scheme. The pixels measured by FDTR are recorded in a random sampling matrix **Π** with bool values, whose element is set to 1 if the location is measured, and the rest of the matrix elements are set to 0. (b) Raw image of interface thermal conductance obtained by the random down-sampling scheme, with only 25% pixels are experimentally measured. (c) Comparison between the original DCT coefficients of the ground truth and the reconstructed coefficients. (d) The reconstructed image of interface thermal conductance.

## B. Implementation of local FDTR measurements

Local thermal property measurements are performed by FDTR, as shown in Figure 2(a). Our FDTR system comprises a 488 nm pump laser and a 532 nm probe laser. Both beams are focused via a 10X objective lens. The root-mean-square beam radius $r_{RMS} = \sqrt{(r_s^2 + r_p^2)/2}$ is measured as 3 μm through beam-offset profiling of the in-phase signal at a high modulation frequency of 50 MHz, in which case the effect of in-plane heat spreading is negligible [33]. A Physik Instrument ® E-727 digital piezo controller is used to precisely control the location of local measurements with an error margin less than 100 nm in the in-plane coordinates $(x, y)$ [34]. A lock-in amplifier (Zurich Instruments ® UHFLI) is used to generate the periodic signal modulating the pump laser and amplify the thermoreflectance signal. In FDTR measurements, the local thermal response is calculated as:

$$H(\omega) = \beta \int_0^\infty \mathcal{G}(\xi, \omega) \exp\left[-\frac{\xi^2 r_{RMS}^2}{4}\right] \xi d\xi, \tag{1}$$

where $\beta$ is a proportionality coefficient determined by the thermoreflectance coefficient, the gain of detectors, and laser power; $\xi$ is the Hankel transform variable (i.e. wave number of a cylindrical thermal wave). $\mathcal{G}(\xi, \omega)$ is Green's function under pulsed point heating in the frequency-Hankel domain, which can be solved from the heat transfer matrix method [17,35]. The amplitude and the phase signals are defined as:

$$H(\omega) = A(\omega) e^{i\phi(\omega)}. \tag{2}$$

Phase signals $\phi(\omega)$ are used to extract the local thermal properties since $\phi(\omega)$ is independent of the laser power in the linear thermoreflectance regime. Further details on thermal models in FDTR are available in Refs [36,37]. To increase the speed of local measurements, the modulation frequencies are superimposed onto the modulation signal, such that thermal phases $\phi$ at different frequencies can be measured at once, as illustrated in Figure 2(b). At each point $(x, y)$ on the sample, the thermal phase $\phi(\omega, x, y)$ is fitted by the thermal model to extract the local thermal property $F(x, y)$, where $F$ represents the thermal property to be measured, such as the thermal conductivity or interface thermal conductance. The thermal property imaging can be collected and represented by a matrix $\boldsymbol{F}$, with the components $F_{mn} = F(x_m, y_n)$ representing the property at the location $x_m = m\Delta x$ and $y_n = n\Delta y$, where $m, n$ are integers counting from zero and $\Delta x, \Delta y$ are the minimal translation steps, set as 1 μm throughout this article. In the CS-FDTR, the original full imaging $\boldsymbol{F}$ was down-sampled as:

$$\boldsymbol{T} = \boldsymbol{\Pi} \circ \boldsymbol{F}, \tag{3}$$

where $\circ$ represents the Hadamard product that multiplies the matrix components correspondingly: $T_{ab} = \Pi_{ab} F_{ab}$. The down-sampled image $\boldsymbol{T}$ contains nonzero components $T_{pq} \neq 0$ only if the location $(p\Delta x, q\Delta y)$ is measured by FDTR, see Figure 1(b).

Although the scanning step sizes $(\Delta x, \Delta y)$ can be much smaller than the spot size $r_{RMS}$ thanks to the high precision of the piezo controller, this does not mean that the spatial resolution of the FDTR image can be taken as the scanning step. In each local measurement, the thermal response is weighted and integrated over the Gaussian profile of pump and probe lasers, and the thermal properties within the beam region are considered uniform. If the sampling interval is smaller than the RMS radius, the laser beams of successive measurements would have a large overlapped area, and the thermal conductivity variations cannot be distinguished. As shown in Figure S1 of Supplemental Material [38], the superimposed Gaussian profile of two successive local measurements shows a single peak when the scanning step is smaller than $r_{RMS}$, and the two peaks representing two measurement locations can only be distinguishable when the scanning step is larger than $r_{RMS}$. Therefore, we use $r_{RMS}$ to estimate the spatial resolution of thermal property imaging.

In the next section II.C, we will describe methods of reconstructing the original image $\boldsymbol{F}$ from the down-sampled image $\boldsymbol{T}$ using $\mathcal{L}_1$-regularized minimization.

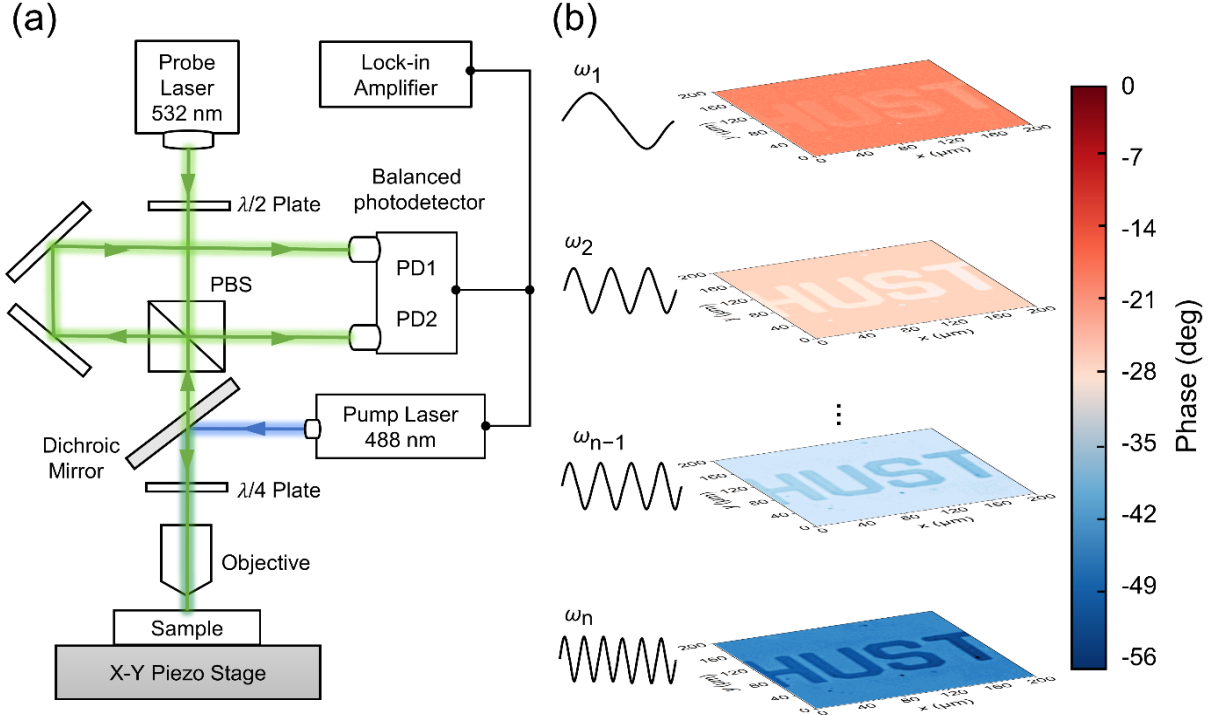

**Figure 2. Thermal property imaging using FDTR.** (a) Schematic of the FDTR experimental setup. (b) Phase images captured at different modulation frequencies.

### C. Image reconstruction

Reconstructing the thermal property imaging $\boldsymbol{F}$ is enabled by their sparse feature. In this work, we use the two-dimensional discrete cosine transform (DCT) to convert $\boldsymbol{F}$ into the representation domain. To implement the image reconstruction algorithm, we vectorize thermal property imaging $\boldsymbol{F}$ into a one-dimensional vector $\boldsymbol{F}_v$ of length $N$, with $N = WL$ the total number of pixels in the $W \times L$ image, by stacking each column. The vectorized image $\boldsymbol{F}_v$ can be expressed as a linear combination of DCT coefficients $\boldsymbol{s}$ as:

$$F_v = (D \otimes D)s = \Psi s, \tag{4}$$

where $D$ is the matrix of one-dimensional DCT, $\otimes$ represents the Kronecker product and $\Psi = D \otimes D$ denotes the transformation matrix correlating the image $F_v$ with the sparse representation $s$. In general, DCT representations of natural images are sparse [32,39], namely, the number of nonzero components in $s$ is usually much smaller than the number of pixels in $F$, see Figure 2(a).

Next, we need to find a measurement matrix that relates the full image $F_v$ and the nonzero components in the down-sampled image $T$ for easy computations using linear algebra. To represent these relations in the form of multiplications, the nonzero components in $T$ are also stored in a vector $F_s$ with the length $M$ equal to the number of local FDTR measurements. $F_s$ and $F_v$ are then related through a $M \times N$ measurement matrix $\Phi$:

$$F_s = \Phi F_v. \tag{5}$$

The measurement matrix $\Phi$ is related to the random sampling matrix $\Pi$ as follows. If $M$ random locations on the sample surface are going to be measured, a series of random integer pairs $(p_j, q_j)$ are first generated with $j \in [0, M)$, $p_j \in [0, W)$ and $q_j \in [0, L)$, and FDTR measurements are performed at $(x_j, y_j) = (p_j \Delta x, q_j \Delta y)$. Correspondingly, the sampling matrix component $\Pi_{p_j, q_j} = 1$. Then the components of the $j$-th measurement $\Phi_{j, Wp_j + q_j} = 1$ in the measurement matrix $\Phi$, and the rest of the components are set to zero, as shown in Figure 3(b). To reconstruct the original $F$, one needs to find the optimal sparse coefficients $s$ that can reproduce the down-sampled signal $F_s$. Combining Eqs. (4-5), the reconstruction problem is therefore solving the following equation:

$$F_s = \Phi \Psi s = \Theta s. \tag{6}$$

Directly solving $s$ from Eq. (6), however, is an ill-posed problem, since the matrix $\Theta = \Phi\Psi$ does not have a full rank, as shown in Figure 3(c).

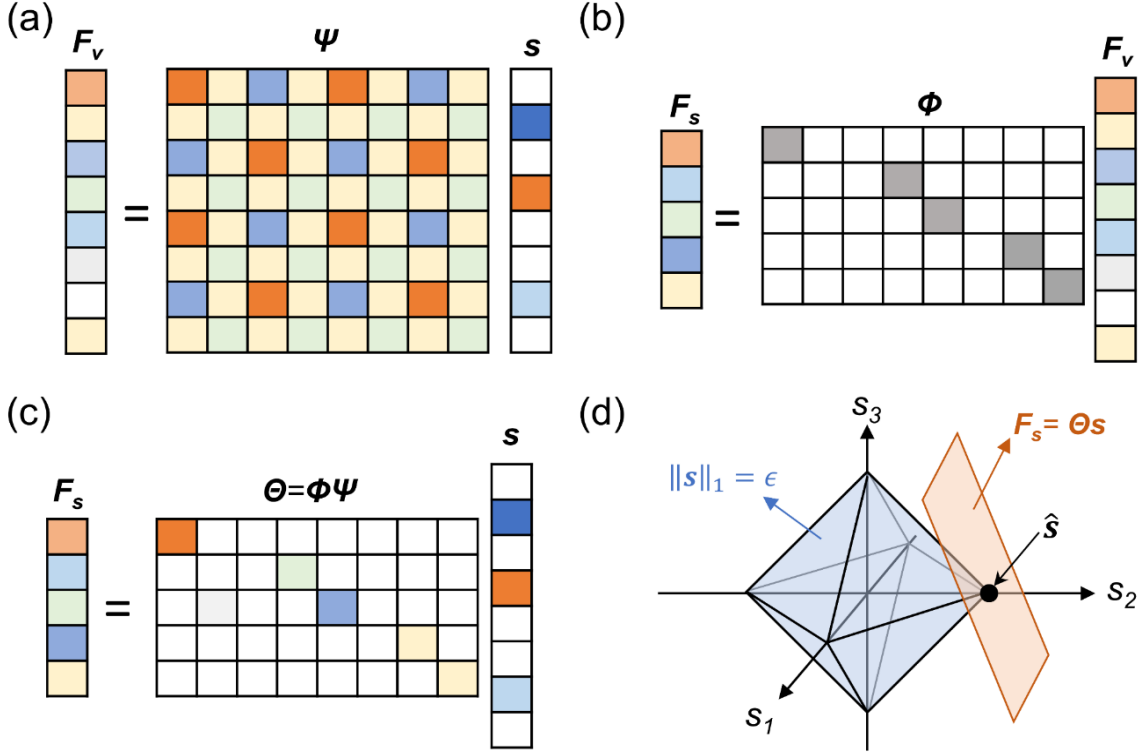

**Figure 3. Image reconstruction using $\mathcal{L}_1$-regularized minimization.** (a) Relation between the flattened image vector $F_v$ and the sparse representation vector $s$. (b) The down-sampled measurement matrix $\Phi$ relating the full image $F_v$ and the down-sampled image $F_s$. (c) The linear problem of finding sparse coefficients $s$ that can reproduce the down-sampled image $F_s$. (d) Illustration of finding the optimal sparse coefficient using $\mathcal{L}_1$-regularized minimization. Denoting $\min\|s\|_1 = \epsilon$, the optimal coefficient $\hat{s}$ is found when the hyperplane $F_s = \Theta s$ intersects with the $\mathcal{L}_1$-hypersphere $\|s\|_1 = \epsilon$.

Generally, the measurement matrix $\Phi$ of random sampling satisfies the restricted isometry property [31,32], which allows solving a sparse $s$ deterministically by minimizing:

$$\hat{s} = \text{argmin}\|s\|_1, \qquad \text{subject to } F_s = \Theta s. \qquad (7)$$

Here, $\text{argmin} f(x)$ means the value of argument $x$ at the minima of $f$. Therefore, $\hat{s}$ denotes the optimal sparse coefficients with the minimal $\mathcal{L}_1$ norm, denoted as $\|s\|_1$. Such a constrained optimization problem is visualized in Figure 3(d). The constraint $F_s = \Theta s$ corresponds to a hyper-plane in the Fourier-domain $s$, and the surface with minimized $\|s\|_1$ is an $\mathcal{L}_1$-hypersphere with vertices resting on the axes in the Fourier domain. The optimal sparse coefficients $\hat{s}$ are found when the hyperplane intersects with the $\mathcal{L}_1$-hypersphere. The geometry of the $\mathcal{L}_1$-hypersphere almost guarantees that $\hat{s}$ resides at certain axes with other components equal to zero, which ensures $\hat{s}$ to be sparse.

To facilitate numerical convergence, the LASSO method [40] can be used to solve the $\mathcal{L}_1$-optimization problem in Eq. (7), by minimizing the following loss function:

$$f(s) = \|F_s - \Theta s\|_2^2 + \lambda\|s\|_1 , \qquad (8)$$

where $\lambda$ is the regularization parameter. $\lambda$ controls the balance between the accuracy and the sparsity of the reconstructed image. In the limit of $\lambda = 0$, minimizing $f(s)$ is reduced to the least-squared minimization problem, and in the limit of $\lambda \gg 1$, the minimization process is seeking the sparsest representation of the image regardless of the reconstruction error. Therefore, proper choice of $\lambda$ is critical. In Eq. (8), the first $\mathcal{L}_2$-term scales with squared value of the property image, and the second term $\lambda\|s\|_1$ must have a similar dimension. As a result, $\lambda$ must have the same dimension of the property image, namely, $\lambda \propto \langle F_s \rangle$, where $\langle F_s \rangle$ is the mean value of the down-sampled image. In this work, $\lambda$ is set as $0.005\langle F_s \rangle$, which has reasonable reconstruction accuracy, as we shall see later for the reconstruction results in Section III. In Figure S2 of Supplemental Material [38], we have also tested different choices of $\lambda$ from $0.001\langle F_s \rangle$ to $0.01\langle F_s \rangle$, all showing robust reconstruction accuracy.

The minimization of $f(s)$ is performed iteratively using the Orthant-Wise Limited-memory Quasi-Newton (OWL-QN) algorithm implemented in the PyLBFGS package [41,42]. After the optimal sparse coefficients $\hat{s}$ are obtained, the image vector is reconstructed by:

$$\widehat{F_v} = \Psi\hat{s}. \tag{9}$$

The flattened vector $\widehat{F_v}$ can be easily reshaped to the two-dimensional image $\widehat{F}$. Finally, the normalized root-mean-squared error (NRMSE) is defined to evaluate the fidelity of the reconstructed thermal property image $\widehat{F}$:

$$u(\widehat{F}, F) = \left[\frac{\sum_{i=1}^{L}\sum_{j=1}^{W}(\widehat{F}_{ij} - F_{ij})^2}{\sum_{i=1}^{L}\sum_{j=1}^{W}(F_{ij})^2}\right]^{1/2}. \tag{10}$$

The NRMSE is the key parameter for determining the appropriate sampling fraction that can achieve fast imaging speed and high quality at the same time. In this work, we characterized the dependence of NRMSE on sampling fraction through the following iterative manner. The thermal property profile is first imaged under a low sampling fraction (5%), and then the rest of the unsampled locations are randomly selected to increase the sampling fraction by 5%. Such a process is repeated until all locations inside the scanning window have been measured.

## III. RESULTS OF THERMAL PROPERTY IMAGING

To demonstrate the feasibility of CS-FDTR, the developed method is applied to image three samples with different thermal property distributions. In Section III.A, CS-FDTR is first applied to map the interface thermal conductance across a heterogeneous interface with an artificially fabricated "HUST" pattern. Next in Section III.B, irregular in-plane thermal conductivity variations of an annealed pyrolytic graphite (APG) sample are measured. Finally, we image the sharp change in thermal conductivity profile across a vertical aluminum-graphite interface in Section III.C.

## A. Mapping thermal conductance across a patterned interface

First, we use a patterned interface with an artificially fabricated "HUST" letter pattern to test the feasibility and performance of CS-FDTR. As shown in Figure 4(a), the patterned interface is prepared by depositing a 6 nm film of chromium (Cr) with the HUST pattern on a silicon substrate using photolithography and a lift-off process. After the patterning, the Cr film was exposed in air to oxidize, in order to increase the contrast ratio in interface conductance within the letter pattern and the rest of the sample surface. Following the lift-off process, the entire sample was coated with a 100 nm thick Au film through sputtering. Figure 4(b) shows the ground truth image measured by point-by-point FDTR scanning, with a sampling interval of 1 μm. The average interface thermal conductance is 30 MW $m^{-2}$ $K^{-1}$ in the region without Cr transition layer, and a decreased thermal conductance of 21 MW $m^{-2}$ $K^{-1}$ is observed inside the region with Au/Cr($Cr_2O_3$)/Si interface, which is probably due to the existence of the native oxide layer. Representative phase signals in Au/Si and Au/Cr($Cr_2O_3$)/Si interfaces are shown in Figure S3(a) of Supplemental Material [38]. Regions with lower interface thermal conductance align well with the as-fabricated Cr pattern. The same region is randomly down-sampled and then reconstructed using the LASSO algorithm. To quantitatively evaluate the imaging quality, we compare the NRMSE between the reconstructed image and the original point-by-point full image at varied sampling fractions in Figure 4(c). Representative raw images and reconstructed images at 5%, 15%, and 25% sampling fractions are shown in Figure 4(d) and Figure 4(e). At the very low 5% sampling fraction, the reconstructed pattern shows considerable distortion. For example, discontinuities are observed in both the reconstructed letter "S" and "H". However, the "HUST" pattern is successfully captured when the sampling fraction is increased to 15%, and with a low NRSME of ~ 5%. The reconstructed thermal conductance image shows clearer edges of the HUST pattern, with a low NRSME of ~ 4% when the sampling fraction is further increased to 25%.

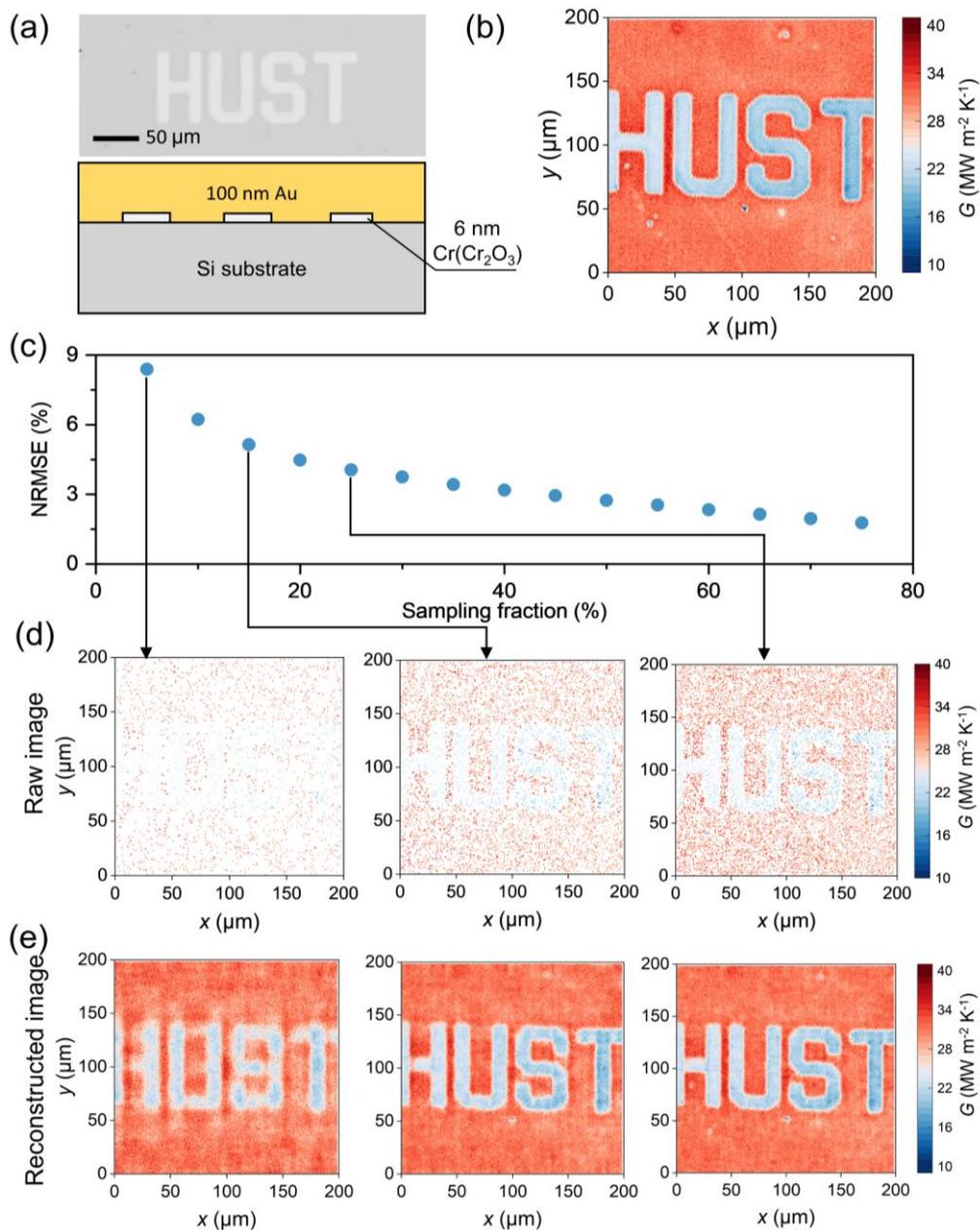

**Figure 4. Imaging thermal conductance across a patterned interface.** (a) Optical image and the cross-section structure of the Au/Cr($Cr_2O_3$)/Si interface with an "HUST" pattern. (b) Ground truth of interface thermal conductance through point-by-point scanning FDTR. (c) Normalized root-mean-squared error (NRMSE) of the reconstructed image at different sampling fractions. (d) representative raw images and (e) reconstructed images at 5%, 15%, and 25% sampling fractions.

**B. Imaging the thermal conductivity variations of annealed pyrolytic graphite**

In addition to the thermal property profiles of artificially patterned samples, we test the capability of CS-FDTR in imaging irregular thermal conductivity variations of an APG sample. We have prepared an APG sample using the chemical vapor deposition method [43], followed by an annealing process at 2800 °C under a uniaxial pressure of ~ 5 MPa to align the graphite layers [44]. The optical microscope image of the APG surface as shown in Figure 5(a) indicates high sample quality with a low density of surface wrinkles. Figure 5(b) shows the in-plane thermal conductivity distribution in an $80 \times 80$ μm$^2$ area measured by point-by-point scanning. The FDTR implemented in this work is primarily sensitive to in-plane thermal conductivity due to the small root-mean-square laser spot radius of 3 μm. To reduce the unknown fitting parameters, the cross-plane thermal conductivity of the same APG sample is separately measured using TDTR with a large spot size of 16 μm (Supplemental Material [38] Figure S4(a)). The in-plane thermal conductivity of APG shows nonuniform distributions with patterns nonvisible in the optical image. Several regions with high thermal conductivity of ~ 1800 W m$^{-1}$ K$^{-1}$ are observed, while most of the regions have a slightly lower thermal conductivity of 1400 m$^{-1}$ K$^{-1}$. As shown in Figure 5(c-e), the CS-FDTR also shows excellent capability of imaging irregular thermal conductivity variations. Even with a sampling fraction down to 10%, the CS-FDTR can successfully capture the shape of high-thermal conductivity (> 1400 W m$^{-1}$ K$^{-1}$) regions, and finer variations inside these high thermal conductivity areas can be further restored when the sampling fraction is increased to above 25%.

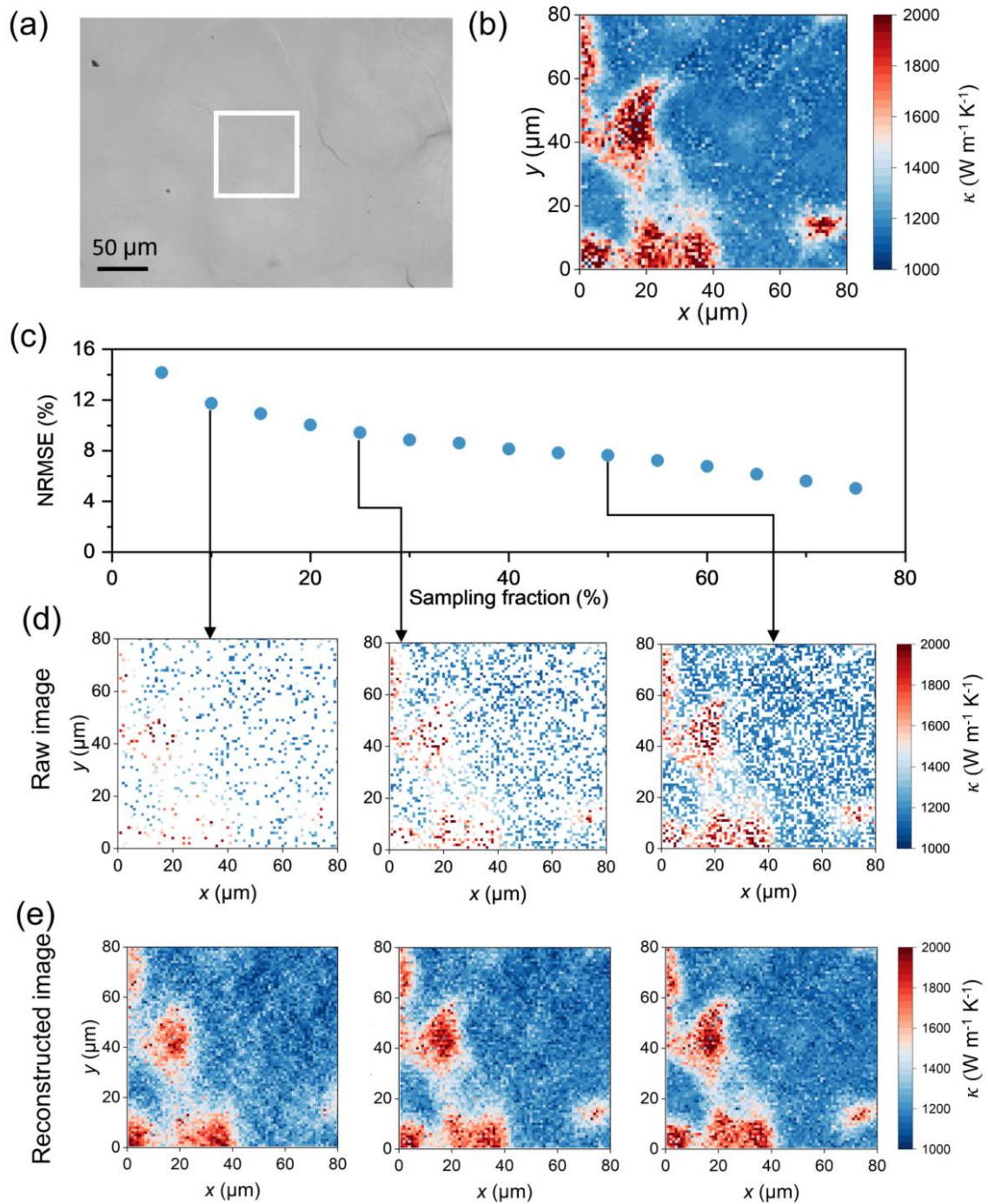

**Figure 5. Imaging thermal conductivity variations of an APG sample.** (a) Optical image of the APG, with the imaging window indicated by the square. (b) Ground truth of the thermal conductivity image of APG using point-by-point full sampling. (c) NRMSE of the reconstructed image at different sampling fractions. (d) Representative raw images and (e) reconstructed images at 10%, 25%, and 50% sampling fractions.

## C. Thermal conductivity profiling near a vertical Al/graphite interface

This part testifies the performance of CS-FDTR in profiling sharp changes in thermal conductivity across an Al/graphite interface perpendicular to the sample surface. The vertical Al/graphite interface is prepared by hot pressing the aluminum and highly ordered pyrolytic graphite together under 50 MPa and 900 K and kept for 30 minutes. The graphite layers are stacked parallel to the vertical interface. Figure 6(a) shows the optical image of the cross-section near the vertical Al/graphite interface. In this configuration with graphite basal planes aligned vertically, FDTR is dominantly sensitive to the thermal conductivity normal to the exposed sample surface (Figure S5(b) of Supplemental Material [38]). The thermal conductivity perpendicular to the sample surface is measured as ~ 1900 W m$^{-1}$ K$^{-1}$ on the graphite side, and the thermal conductivity of the aluminum is measured as ~ 150 W m$^{-1}$ K$^{-1}$ (Figure S5(a) of Supplemental Material [38]). The full thermal conductivity image measured using point-by-point scanning is shown in Figure 6(b). Figure 6(c) further shows that the NRMSE of the Al/graphite sample shows a moderate decrease as the sampling fraction increases. Representative raw images and reconstructed images at 25%, 50%, and 75% sampling fractions are shown in Figure 6(d) and Figure 6(e). The representative line profiles of both the ground truth (full sampling) and reconstructed thermal conductivity images near the Al/graphite vertical interface are shown in Figure S6 of the Supplemental Material [38]. At the sampling fraction of 25%, the shape of the Al/graphite boundary is restored, but the reconstructed thermal conductivity map is quite noisy on the Al side. Even at a 50% sampling fraction, the NRMSE is still ~ 15%, larger than those of the previous two samples. This is because the sharp change in thermal conductivity near the interface will result in many high-spatial-frequency components in the Fourier domain (Figure S7 of Supplemental Material [38]). As a result, the sparsity of the thermal conductivity image near the interface is not as pronounced, requiring higher sampling fractions (~50%) for accurate reconstruction compared with the previous two samples. In the following section, we will further analyze why a higher sampling fraction is needed for this sample.

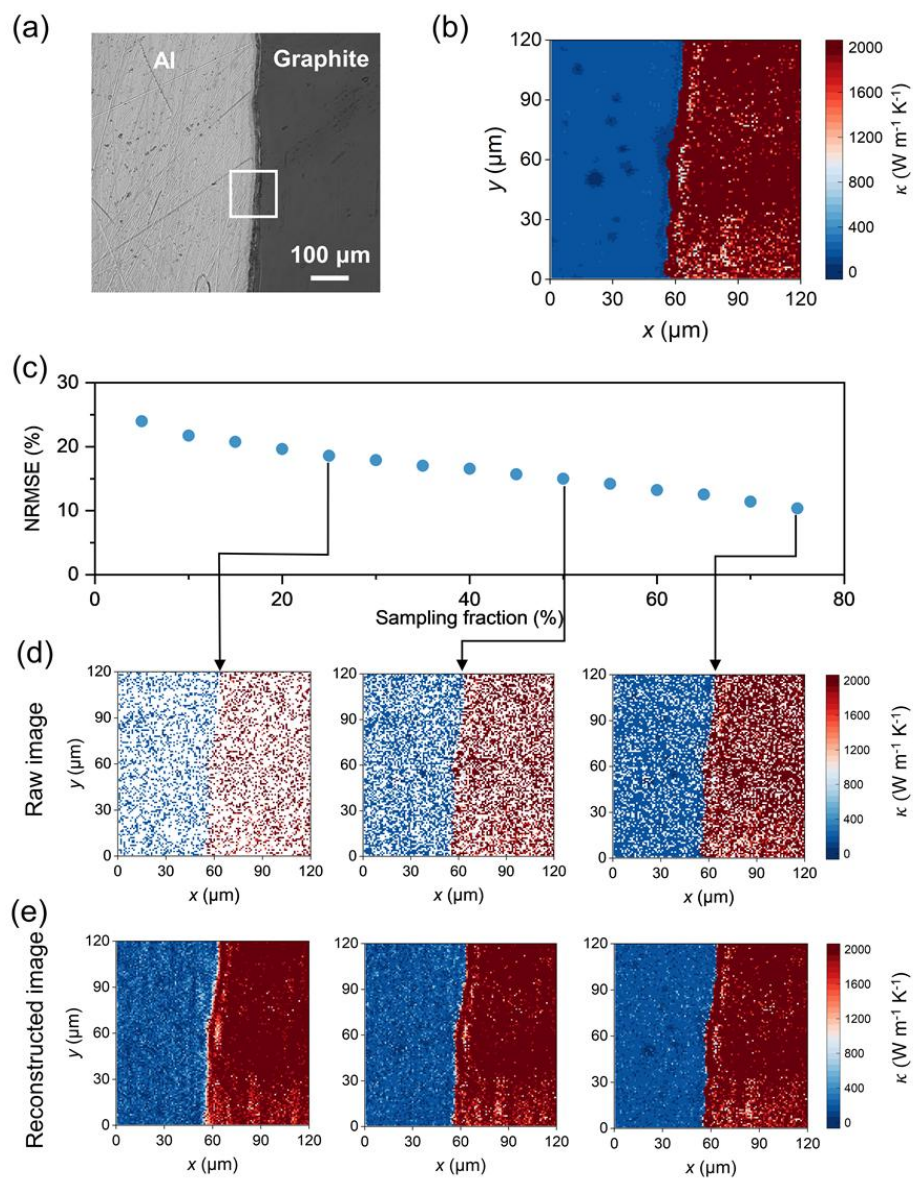

**Figure 6. Imaging sharp thermal conductivity changes across a vertical Al/graphite interface.** (a) Optical image of the Al/graphite composite, with the imaging window indicated by the square. (b) The ground truth image of thermal conductivity near the Al/graphite interface. (c) NRMSE of the reconstructed image at different sampling fractions. (d) representative raw images and (e) reconstructed images at 10%, 25%, and 50% sampling fractions.

# IV. DISCUSSION ON SPARSITY AND RECONSTRUCTION ACCURACY

To provide insights into why a higher sampling fraction is required to obtain a reasonable accuracy for certain cases, we performed a quantitative analysis of the DCT coefficients for the three samples aforementioned. Since we are concerned with spatial variations in thermal properties, all DCT coefficients are normalized by the mean value of the image $\langle \boldsymbol{F} \rangle$. Figure 7(a) shows the typical distribution of the original coefficients $\boldsymbol{s}_{true}$ of the ground truth and the reconstructed coefficients $\hat{\boldsymbol{s}}$ through $\mathcal{L}_1$-minimization, for the image of interface conductance with "HUST" patterns. Similar plots for other samples are shown in Figure S8 of the Supplemental Material [38]. CS-FDTR shows excellent performance in reconstructing large coefficients, while the distribution of coefficients with magnitudes smaller than 0.1 shows deviations from that of the ground truth image. This is also manifested in Figure 7(b) showing the relative deviation $|\hat{\boldsymbol{s}} - \boldsymbol{s}_{true}| / |\boldsymbol{s}_{true}|$ of each DCT coefficient, where the large coefficients are generally reconstructed with a small relative deviation. Due to the $\mathcal{L}_1$ regularization term $\lambda \|\boldsymbol{s}\|_1$ in Eq. (8) that favors the highest sparsity of $\hat{\boldsymbol{s}}$, there will be a considerable number of small coefficients in $\boldsymbol{s}_{true}$ reconstructed as zeros in $\hat{\boldsymbol{s}}$. This is manifested as the dots distributed horizontally with a relative deviation of unity, see Figure 7(b). We can define a reconstruction edge $s_{edge}$ as the maximum coefficient with a relative deviation equal to one, below which there will be significant reconstruction errors. With the increasing sampling fraction, the reconstruction edge $s_{edge}$ decreases, this means that more DCT coefficients are accurately reconstructed in CS-FDTR, resulting in decreased NRMSE at higher sampling fractions.

This trend of decreased $s_{edge}$ with increasing sampling fraction is consistent among the three samples we measured, as shown in Figure 7(c). However, the reconstruction edges for the three samples are decreasing in the following order: patterned interface > APG > Al/graphite, at a fixed sampling fraction. This can be understood by analyzing the intrinsic sparsity of the property image. For the HUST-patterned interface, most large coefficients are concentrated in the region with small spatial frequencies, and the majority of the components in $s$ are negligibly small; in contrast, the sharp interface of Al/graphite has much more nonnegligible components with higher spatial frequencies (Figure S7 of Supplemental Material [38]). Namely, the image of the HUST-patterned interface is sparser than the image of the Al/graphite sample. We can quantify the image sparsity as the ratio between the number of components in $s$ below $0.01\langle F \rangle$ and the total number of elements in $s$. It is clearly shown in Figure 7(c) that a sparser image corresponds to a lower reconstruction edge, when the same sampling fraction is used in CS-FDTR.

As shown in Figure 7(d), the smaller the reconstruction edge $s_{edge}$, the lower NRMSE (*i.e.* better agreement) between the reconstructed image and the ground truth. This behavior can be qualitatively captured using a brutal stepwise cut-off simulation in the relative deviation of DCT coefficients. Below $s_{edge}$, all relative deviations are regarded as one, while the coefficients larger than $s_{edge}$ are assumed to be exactly reconstructed. Such an assumption corresponds to the sparsest limit for coefficients below $s_{edge}$ and the ideal reconstruction limit for large coefficients above $s_{edge}$. For the patterned interface and the APG sample with higher sparsity values, the simulated NRMSEs (lines in Figure 7(d)) agree well with the practical reconstruction (symbols in Figure 7(d)). However, for the Al/graphite composite with very low sparsity (3.48%), the NRMSE deviates from our simulation, because the sharp thermal conductivity change across the interface is embedded into lots of small and high-frequency components, which can only be effectively captured at a higher sampling fraction to reduce the reconstruction edge $s_{edge}$.

Based on the above analysis, we now provide suggestions on how to select a reasonable sampling fraction using CS-FDTR. For composite materials, interfaces, and patterned structures with clustering of similar properties in certain regions, thermal property images of these samples are usually sparse in the Fourier domain. Solving the $\mathcal{L}_1$-regularized minimization problem can typically result in a reconstruction edge $s_{edge}$ on the same order of magnitude as the mean value of the image (Figure 7(d)), which in turn corresponds to a sampling fraction on the order of 10% (Figure 7(c)). Practical CS-FDTR measurements on the HUST-patterned interface and APG samples show that a 25% sampling fraction is enough to obtain excellent imaging quality for these sparse structures. However, for sharp property changes across an isolated interface, the image is much less sparse, resulting in a large reconstruction edge $s_{edge}$. Therefore, lots of small coefficients cannot be reconstructed accurately in the $\mathcal{L}_1$-regularized minimization, and the NRMSE can only be effectively reduced with more pixels measured experimentally at higher sampling fractions up to 50%.

In practice, the sampling fraction can also be determined by an iterative process of random sampling and reconstruction. For example, one can first measure the thermal property under a low sampling fraction and perform the reconstruction. If a finer imaging resolution is required, additional random measurements can be performed at the rest unmeasured locations, and the image reconstruction can be repeated. This sampling-reconstruction process can be iterated until satisfactory imaging quality is achieved.

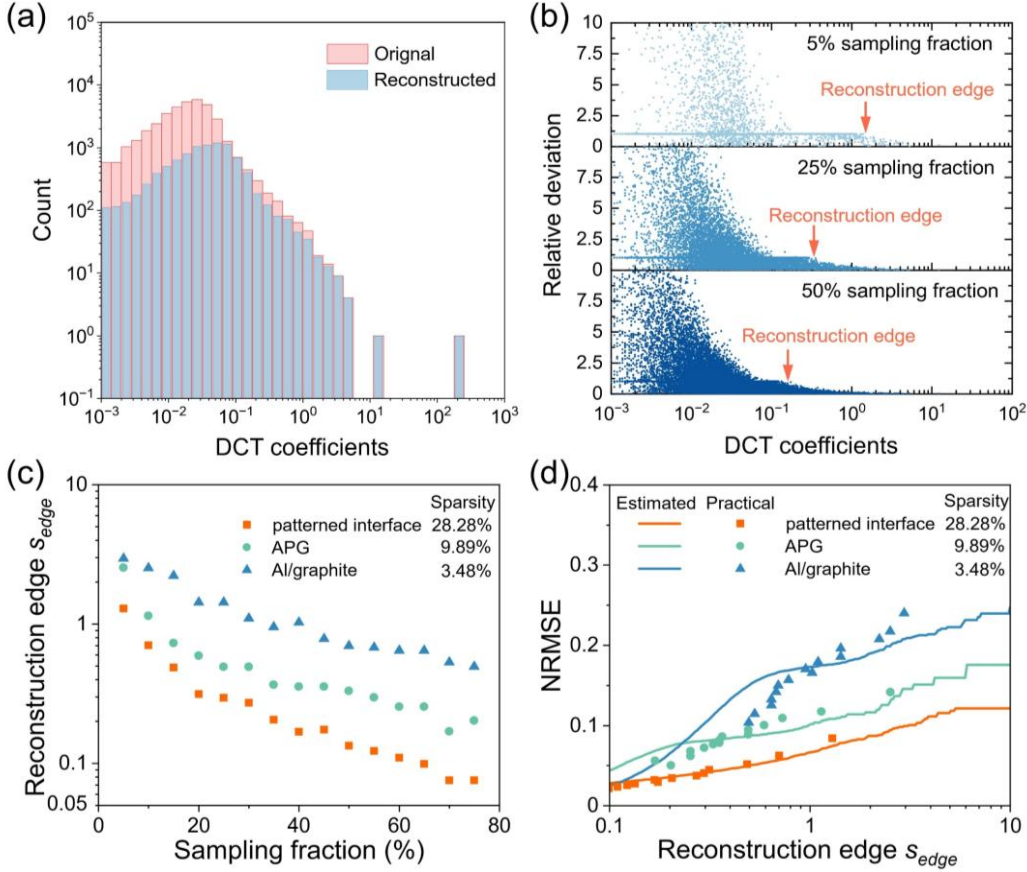

**Figure 7. Effect of image sparsity on reconstruction accuracy using CS-FDTR.** (a) The histogram of DCT coefficients for the original and the reconstructed image of the interface conductance with "HUST" patterns at 25% sampling fraction. (b) The relative deviation of the reconstructed coefficients $\hat{s}$ with respect to the coefficients $s_{true}$ of the ground truth: $|\hat{s} - s_{true}|/s_{true}$. The reconstruction edge is the threshold below which reconstructed DCT coefficients have significant relative error. All DCT coefficients in (a) and (b) are normalized by the mean value of the image. (c) Relation between the reconstruction edge and the sampling fraction for the three images with different sparsity. (d) Estimated and practical NRMSE of the three images with different sparsity.

## V. SUMMARY

In this study, we implemented the CS-FDTR that enables high-throughput thermal property imaging by taking sparse random sampling measurements and reconstructing the property maps through $\mathcal{L}_1$-regularized minimization. The feasibility of the proposed CS-FDTR imaging scheme is validated by measuring the thermal conductance of a patterned interface, irregular thermal conductivity variations of an annealed pyrolytic graphite sample, and thermal conductivity near a vertical Al/graphite interface at different sampling fractions. For the former two samples, excellent imaging fidelity quality is achieved with low sampling fractions of 15% and 25%, with NRSMEs below 10%. For imaging sharp variations of thermal conductivity near an Al/graphite interface, a higher sampling fraction of 50% is needed for an NRSME of 15%, due to the lower imaging sparsity. Our work simultaneously reduces the signal acquisition and data processing time without sacrificing the imaging quality, and enables high-throughput thermal characterization essential for materials discovery, screening, and optimization.

## Supplemental Material

See the supplementary material for more details more information about the spatial resolution of FDTR images, selection of the regularization parameter, typical FDTR signal and sensitivity analysis, line profile of thermal conductivity distributions, and sparsity analysis.

## Acknowledgments

X.Q. acknowledges the support from the National Key R & D Project from the Ministry of Science and Technology of China (Grant No. 2022YFA1203100). T.H.L acknowledges support from the National Natural Science Foundation of China (NSFC, Grant No. 52076089)

## Conflict of Interest

The authors have no conflicts to disclose.

## Data Availability Statement

The data that support the findings of this study are available from the corresponding author upon reasonable request.